\documentclass[twocolumn,showpacs,preprintnumbers,amsmath,amssymb,apsrev]{revtex4}
\usepackage{graphicx}% Include figure files
\usepackage{dcolumn}% Align table columns on decimal point
\usepackage{bm}% bold math

\begin{document}

\preprint{}

\title{Large Force Fluctuations in a Flowing Granular Medium}

\author{Emily Longhi}
\author{Nalini Easwar}%
 \email{neaswar@smith.edu}
\affiliation{%
Dept. of Physics, Smith College, Northampton, MA 01063 U.S.A.}%
\author{Narayanan Menon}
 \email{menon@physics.umass.edu}
\affiliation{
Dept. of Physics, University of Massachusetts, Amherst, MA 01003 U.S.A.}
%\date{\today}

\begin{abstract}
We report the characteristics of the temporal fluctuations in the local force 
delivered to the wall of a 2D hopper by a granular medium flowing through it.  
The forces are predominantly impulsive at all flow rates for which the flow does
not permanently jam. The average impulse delivered to the wall is much larger 
than the momentum acquired by a single particle under gravity between collisions,
reflecting the fact that momentum is transferred to the walls from the bulk of 
the flow by collisions.  At values larger than the average impulse, the 
probability distribution of impulses is broad and decays exponentially 
 on the scale of the average impulse, just as it does in static 
granular media.  At small impulse values, the probability distribution 
evolves continuously with flow velocity but does not show a clear signature 
of the transition from purely collisional flow to intermittently jamming flows.  
However, the time interval between collisions tends to a power law 
distribution, $P(\tau)\sim \tau^{-3/2}$, thus showing a clear dynamical signature of 
the approach to jamming. 
\end{abstract} 


\maketitle

Sand, rather than a liquid, is chosen for the contents of an hourglass because the rate of 
efflux from a column of sand flowing under gravity does not depend on the height to 
which it is filled. Likewise, in a static column of sand, the pressure at a given depth is 
independent of the height of the column above it \cite{Jaeger96}. In both cases, this is because the 
weight of the sand in the interior of the column is borne by the lateral containing walls.  
In the case of a static granular column, experiments \cite{Dantu&Travers,Liu1995,Howell1999}, 
theory \cite{Coppersmith1996,CatesChaos99}, and numerical 
simulations \cite{Liu1995,Radjai1996, Makse2000} have all shown that stresses in the bulk 
are transmitted to the walls in a very spatially inhomogeneous fashion.  This is demonstrated 
by two independent observations: first, the distribution of forces, $P(f)$, at the boundaries of the medium is 
exponential, rather than gaussian \cite{Mueth1998,Miller1996} Secondly, grains that are highly stressed are 
organized into filamentary, linear structures called `force chains' that carry a large 
fraction of the stress \cite{Dantu&Travers,Liu1995,Howell1999}.

In this article we report experiments directed toward producing a complementary 
understanding of a flowing granular medium. The force chains in a static medium are 
unstable to perturbations perpendicular to their length; when jostled by other grains 
moving in a flow, do they merely become short-lived, or do they melt away entirely? 
Concomitant to this presumed annealing away of the stress inhomogeneities, does the 
broad, exponential force distribution in the static case become gaussian when grains 
move and rearrange?  The answers to these questions have broad consequences for the 
prospects of applying continuum theories to flowing granular matter.  If indeed the 
bounding walls of a medium can communicate with the interior by transient force chains, 
then any approach to a continuum limit must take into account these long length scales. 
Previous analyses of the force distributions in a static granular medium 
\cite{Coppersmith1996}, show that an 
exponential at large forces may be obtained from scalar models in which the weight of a 
bead is borne by the beads beneath it in some random proportion.  While force chains do 
not naturally emerge from these models it is tempting to speculate that force chains and 
broad force distributions are related manifestations of the large stress inhomogeneities in 
a granular medium.  This point of view is also suggested by recent simulations that show 
a simultaneous narrowing of force distributions and a blurring of the force chains \cite{Makse2000}, as 
static granular media are subjected to compression.  On the other hand, a recent proposal 
\cite{OHern2001} for a unified description of jamming in thermal as well as non-thermal systems 
identifies signatures of the approach to a jammed state in the force distribution, $P(f)$.  In 
contrast to earlier models, they suggest that loss of mobility is due to the formation of 
force chains, whose presence is most clearly signalled by the scarcity of unstressed 
regions (i.e. a dip or plateau in $P(f)$ at $f < \bar f$, the average force) rather 
than by the exponential tails at large force values.  The formation of a  plateau in 
$P(f)$ has been also been reported in a recent simulation \cite{Silbert2001} of grains 
flowing down an inclined chute.

In this article we report measurements of the temporal fluctuations of the forces delivered 
to the side walls of a hopper (shown in Fig.1a) which contains a 2D flow of steel balls 
with diameter $d$=3.125mm.  The sides of the hopper are at a fixed angle of $10^{\circ}$ to the 
vertical, so that there are no locations within the cell where the flow stagnates. A 
transducer (PCB Piezotronics model 209C01) sits flush against one of the sidewalls of 
the hopper and measures forces normal to that wall. The diameter of the head of the 
transducer $\approx d$; thus there is only one ball against the transducer at any time.  For most of 
the measurements we report, the transducer is located at a constant distance of 5 cm (=16$d$)from 
the opening of the hopper.  The flow velocity through the hopper is varied by changing 
the opening at the bottom of the hopper in several steps from 3$d$ to 16$d$. Video 
measurements of $V_f$, the average flow velocity at the location of the transducer, were in accord 
with expected bulk features of the flow.  The efflux rate was a constant during the flow at 
each opening, and showed a power law dependence on the size of the opening, with  
an exponent of 1.5.  Previous experiments in hopper flow have shown a rich variety of 
dynamics including density waves \cite{Baxter1989} and regimes of intermittent flow 
and shocks \cite{Veje1996-97}.  These more complex phenomena do not appear in our 
experimental configuration. Our measurements are similar to those in Ref. 15 
however, in those experiments the hopper angle is $45^{\circ}$ and the flow is 
frictional rather than collisional.
 
%--------------------------------------------------------------------------------------------

\begin{figure}[t]
\includegraphics[width=.5\textwidth]{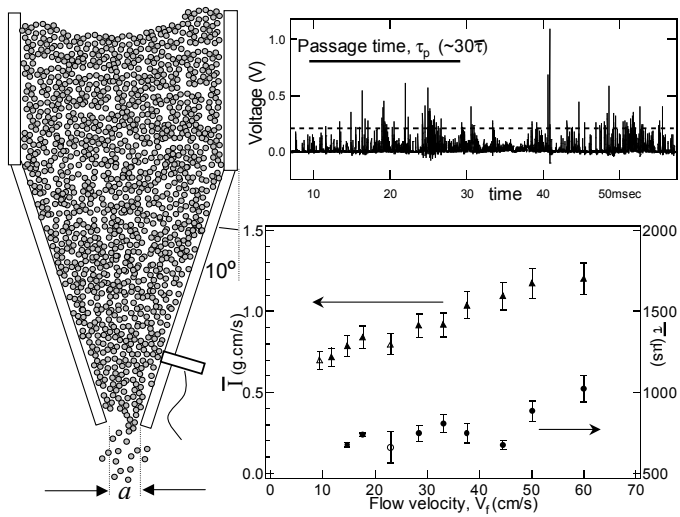}% various averages
\caption{\label{average} (a) Sketch of the experimental set-up. (b) A section of the voltage trace as a 
function of time during a drain. $\tau _{p}$ is the average passage time for a ball across the 
transducer and $\bar \tau$ is the average time interval between peaks. 
The dotted line represents the average height of the peaks in the trace. (c) Average 
impulse $\bar I$ (triangles) and average time interval  $\bar \tau$  (circles) as a function of flow velocity, $V_f$ 
at the transducer.}
\end{figure}

%-------------------------------------------------------------------------------------------
The voltage signal from the transducer was sampled at a rate of 100 kHz with 12-bit 
resolution.  Fig 1b shows a section of a typical trace of voltage as a function of time, 
obtained during a drain. The trace represents the passage of several balls; since the flow 
is very dense there is no evident signature of individual balls.  However, the trace is made 
up of clearly separated peaks indicating that the forces against the walls are collisional 
despite the fact that particles visually appear to be in contact At the two lowest flow 
velocities we report (opening sizes of $3d$, and $3.3d$), occasional sustained contacts begin 
to appear in the force trace though the flow does not jam permanently.  However, at all 
other flow velocities, there are three well-separated time scales as shown in Fig. 1b: The 
average passage time of a ball past the transducer head, $\tau_{p} (=d/V_{f})$, is much greater than 
the average time between collisions, $\bar \tau$, which in turn is much greater than the width of 
a peak.  Thus a ball makes several distinct collisions with the transducer as it flows past 
its face.  
%--------------------------------------------------------------------------------
\begin{figure}[b]
\includegraphics[width=.5\textwidth]{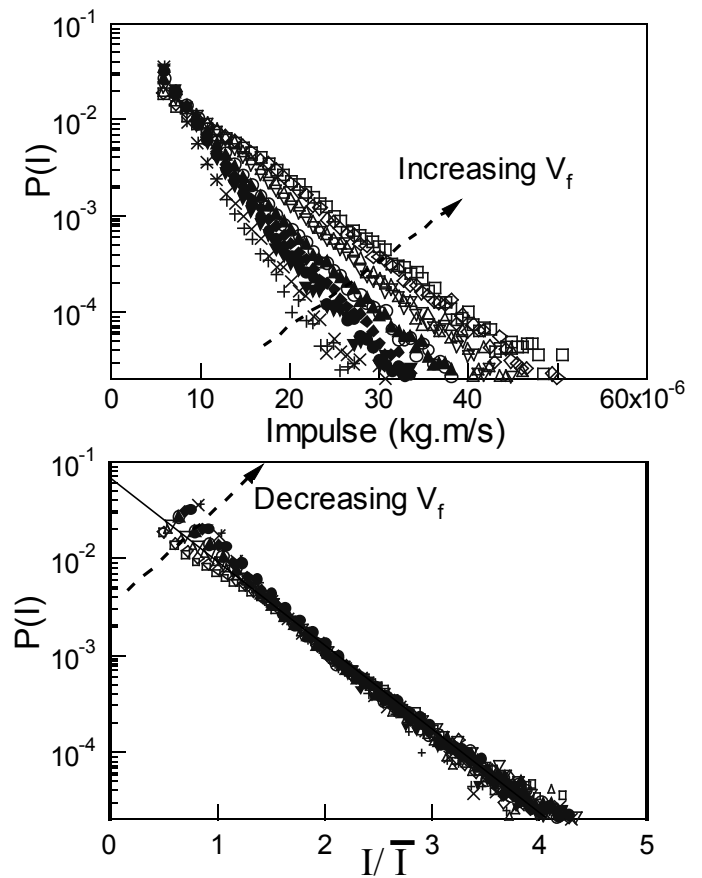}% Impulse histograms
\caption{\label{impulse} (a) Impulse histograms on a log-linear scale for various 
$V_f$ (all in cm/s): $9.4 (X)$, $11.7 (+)$, $14.7 (\bullet)$, $17.7 (\blacklozenge)$,
 $23.0 (\blacktriangledown)$, $28.4 (\blacktriangle)$, $33.1 (\circ)$,
 $37.6 (\bigtriangledown)$, $44.5 (\bigtriangleup)$, $50.1 (\lozenge)$ and $60.0 (\square)$. 
The flow velocities correspond to opening sizes ranging from $a=3d$ to $a=16d$.  
The data at 9.4 cm/s and 23 cm/s correspond to the transducer located higher up in the flow.  
(b) Impulse histograms of Fig.2a scaled to the average impulse $\bar I$ for each flow velocity.}
\end{figure}
%---------------------------------------------------------------------------------
Peaks in the voltage signal above a noise threshold are identified and calibrated against 
known impulses to obtain the normal component of the momentum, $I$, transferred during 
the impact.  The horizontal dashed line in Fig. 1b shows the average peak height 
corresponding to an average impulse, $\bar I = 8\times 10^{-6} kg.m/s.$ Since the wall has to provide 
enough upward momentum to prevent the acceleration of balls along the entire width of 
the channel, it is not surprising that $\bar I$ is greater than the vertical momentum that a ball 
would gain falling in gravity between collisions, which is estimated as $g\bar \tau \approx 1\times 10^{-6} 
kg.m/s$.  In Fig. 1c, we show the variation of the average impulse, $\bar I$, and the average 
collision time $\bar\tau$, as a function of flow velocity, $V_f$.  As might be expected, $\bar I$ 
increases monotonically with flow velocity, $V_f$; $\bar \tau$ also increases modestly with $V_f$.  
In what follows, we discuss the significant features of the distributions of these 
quantities.

In Fig. 2a we plot the probability distribution of the magnitude of impulses 
for several flow rates.  The distributions, $P(I)$, show exponential decay at large $I$, just as 
in the case of static granular media. Furthermore, $P(I)$ retains this exponential shape as 
the opening size is increased from the smallest values that allow sustained flow ($3d$) to 
relatively large openings ($16d$). Fig. 2a also includes data taken at two openings with 
the transducer located further up (10 cm) from the opening where there is much greater 
crystalline ordering of the beads than at the downstream position where most of the other 
data are taken.  As in static granular media \cite{Blair2001}, the exponential force distribution is 
unchanged by the degree of order in the packing of the beads.  Furthermore, the curves in 
Fig. 2a vary monotonically with the flow velocity at the location of the transducer 
irrespective of the location of the transducer.  When the impulse axis is scaled by the 
average impulse for each flow velocity, the large forces ($I >> \bar I$) all collapse and show 
exponential behaviour as shown in Fig. 2b. 

The fact that the broad, exponential distribution of forces seen in the static situation is not 
narrowed by the flow allows for two possibilities.  The first is that some dynamic version 
of force chains persists in flow, with chains dissolving and forming through the bulk of 
the flow, just as they do in the quasistatic shearing regime explored in Ref. 4.  The second 
possibility is that force chains and exponential force distributions are unrelated 
phenomena.  The latter possibility is developed in a recent study of the transition to a 
jammed state where O'Hern et al.\cite{OHern2001} show that in simulations of Lennard-Jones 
potentials as well as in a 2D model of a sheared foam, the distribution of forces at large 
force is dictated only by the interparticle potential at short distances and is not affected by 
shear rate or temperature; for steep enough potentials, an approximately exponential form 
results.  Their simulations indicate that the chief signature of the transition to a jammed 
state (which in the case of granular media is known to have force chains) comes from 
features in the force distribution at low values of force, in particular, a peak or a plateau 
develops close to the average value of force as the temperature or shear rate is lowered 
toward the threshold of jamming.  This is a significant new idea, in that it identifies a 
static quantity –- the force distribution –-  as the  sign of a jamming transition.  Our data do 
show some of the features of the scenario of O'Hern et al. As discussed, the large force 
tail of $P(I)$ is exponential at all flow rates.  Likewise, as seen in Fig. 2b, the shape of $P(I)$ 
at small values of forces changes with flow rate.  However, at slower flow rates, $P(I)$
tends to move upwards rather than bend down and form a peak.  Since we are only able 
to reliably resolve impulses above a threshold value dictated by noise and transducer 
response, we cannot rule out the formation of a peak at extremely low forces, however 
any such peak can only form at force values well below the average, in contrast to the 
examples studied by O'Hern et al.  This comparison is complicated by the fact that we 
measure a time-averaged impulse, whereas the simulations of Refs. 11 and 12 measure 
the instantaneous force.  A further caveat in making the comparison to Ref. 12 is that in 
their geometry, grains interact via enduring, plastic, intergrain contacts just prior to 
jamming.  In our flow geometry, the collisional regime extends almost all the way to the 
jamming threshold \cite{To2001}.

%---------------------------------------------------------------------------
\begin{figure}[b]
\includegraphics[width=.5\textwidth]{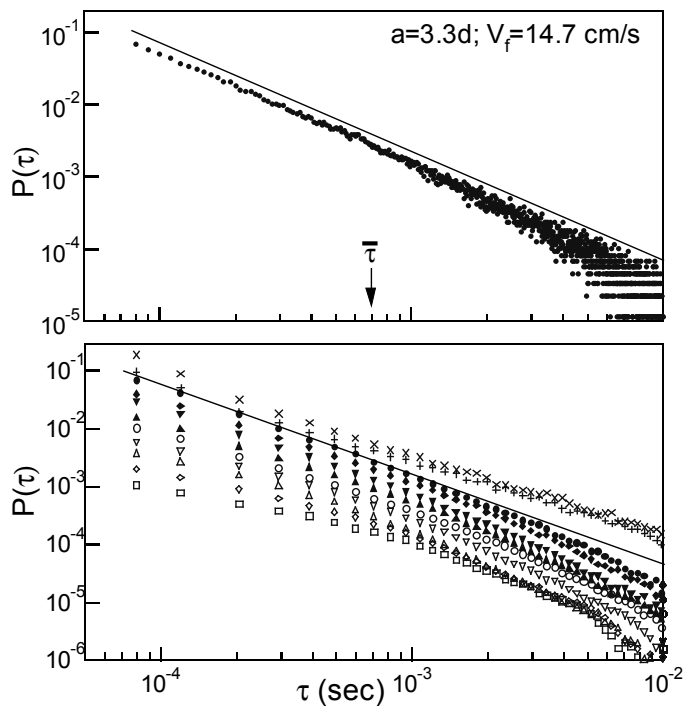}% time intervals
\caption{\label{intervals} (a)Probability distributions, $P(\tau)$, of the time intervals $\tau$
between collisions, on a log-log scale for a flow velocity of $V_f=14.7cm/s$ 
corresponding to $a=3.3d$.  The solid line corresponds to a power law $P(\tau) \sim \tau^{-3/2}$.  The 
average time interval between impulses $\bar \tau$, is marked in the figure.  
(b) $P(\tau)$ on a log-log scale for different flow velocities, $V_f$ (all in cm/s):
 $9.4 (X)$, $11.7 (+)$, $14.7 (\bullet)$, $17.7 (\blacklozenge)$,
 $23.0 (\blacktriangledown)$, $28.4 (\blacktriangle)$, $33.1 (\circ)$, 
$37.6 (\bigtriangledown)$, $44.5 (\bigtriangleup)$, $50.1 (\lozenge)$ and $60.0 (\square)$.
The corresponding opening sizes range from $a=3d$ to $a=16d$.  
The data at 9.4 cm/s and 23.0 cm/s are taken with the transducer located higher up in the flow. 
The curves are displaced vertically for clarity.  The solid line is a power law: $P(\tau) \sim \tau^{-3/2}$}
\end{figure}
%----------------------------------------------------------------------------

Even though we do not observe a static signature of jamming via $P(I)$, we observe a clear 
dynamical signature of the transition to the jamming distribution in, $P(\tau)$, the distribution of the time 
intervals between collisions.  In Figure 3 we show $P(\tau)$ plotted against $\tau$ on a log-log 
scale.  As shown earlier in Fig. 1c, the measured average time interval $\bar \tau$ changes very 
little with $V_f$. However, the distribution $P(\tau)$, gets steadily broader (Fig. 3b) as the flow 
velocity is reduced and tends to a power-law $P(\tau) \sim \tau^{-3/2}$ (Fig. 3a) at a flow velocity of 
$V_f=14.7 cm/s$ corresponding to $a=3.3d$.  In the experiment, there is potentially a long-time 
cut-off for this power law set by the transit time of a ball across the transducer face.  For 
the data in Fig. 3b, this time-scale $\sim 20  msecs$, which lies beyond the largest time interval 
for which we are able to obtain good statistics. However, if the power-law of $\tau^{-3/2}$ is 
indeed the asymptotic shape of the distribution, then the mean time interval $\bar \tau=\int P(\tau)d\tau $ 
tends to diverge just as in a glass transition (even though the average time 
computed from a finite, albeit large, data set shows a relatively innocuous dependence on 
$V_f$ as in Fig. 1c).  At even slower flow velocities (the two curves above the solid line in 
Fig. 3b), there are increasingly frequent instances of sustained contact: at $V_f = 11.7 cm/s$  
and $14.7 cm/s$ the percentage of the time spent in sustained contact is $9 \%$ and $4 \%$, 
respectively, compared to $0.3\%$ at $V_f=44.5 cm/s$.  This leads to ambiguities in defining the 
time interval between collisions possibly resulting in the dashed curves being even 
broader than the power-law of $\tau^{-3/2}$.  A  simulation of gravity-driven channel 
flows \cite{Denniston1999} also showed a power-law distribution of collision times with an exponent of 
close to –3.  In that simulation, however, they were able to compute the distribution of 
collision times for particles in the bulk, averaged over time, whereas in the experiment 
we only have access to the collisions at a fixed location on the wall and cannot keep track 
of any correlations that are convected down the flow with individual balls.

In summary, our data show that the collisional regime extends very close to the threshold 
of jamming.  The broad, exponential distribution of forces in the jammed, static case 
extends right through the jamming transition and up to the largest flow velocities that we 
are able to explore.  The small force end of the force distribution does evolve as the flow 
is slowed down, but the indication of an imminent jammed state appears via a dynamical 
quantity, namely, the distribution of collision times,  $P(\tau)$.  The existence of transient force 
chains in a flowing medium remains unresolved by the measurements reported here, 
though it is clear that if force chains are intimately connected with exponential force 
fluctuations, then they must persist in even the most rapid flows we have investigated. 
We are pursuing this issue further by direct measurement of spatial correlations between 
different points in the flow.

%------------------------------------------------------------------------------------------------------

\begin{acknowledgments}
We thank S. Iwanaga, K. Byers, F. Rouyer, S.M. Dragulin for important contributions to 
the experiment, and S.R. Nagel, A.J. Liu, and O. Narayan for helpful comments.  We also 
gratefully acknowledge support from  NSF CAREER DMR-9874833, the UMass MRSEC (DMR-9809365), 
and the Schultz Foundation.
\end{acknowledgments}

\newpage 
%\bibliography{Klebert_bib}% Produces the bibliography via BibTeX.

\end{document}